\documentclass[lettersize,journal]{IEEEtran}
\usepackage{amsmath,amsfonts}
\usepackage{algorithmic}
\usepackage{algorithm}
\usepackage{array}
\usepackage[caption=false,font=normalsize,labelfont=sf,textfont=sf]{subfig}
\usepackage{textcomp}
\usepackage{stfloats}
\usepackage{url}
\usepackage{verbatim}
\usepackage{graphicx}
\usepackage{cite}

\usepackage{graphicx}
\usepackage{amsmath,amssymb,amsfonts}
\usepackage{diagbox}
\usepackage{booktabs}
\usepackage{multirow}
\usepackage{xcolor}
\begin{document}

\title{Brain Stroke Lesion Segmentation Using Consistent Perception Generative Adversarial Network}

\author{Shuqiang Wang, Zhuo Chen, Senrong You, Bingchuan Wang, Yanyan Shen, Baiying Lei
\thanks{Corresponding Authors: leiby@szu.edu.cn,yy.shen@siat.ac.cn;}
\thanks{Shuqiang Wang, Zhuo Chen, Senrong You, and Yanyan Shen are with 
Shenzhen Institutes of Advanced Technology, Chinese Academy of Sciences,
Shenzhen 518055, China. } 
\thanks{Bingchuan Wang is with School of Automation, Central South University,
Changsha 410083, China } 
\thanks{Baiying Lei is with School of Biomedical Engineering, Shenzhen University,
Shenzhen 518060, China.}
}

\markboth{Journal of \LaTeX\ Class Files,~Vol.~14, No.~8, August~2021}%
{Shell \MakeLowercase{\textit{et al.}}: A Sample Article Using IEEEtran.cls for IEEE Journals}


\maketitle

\begin{abstract}
The state-of-the-art deep learning methods have demonstrated impressive
performance in segmentation tasks.
However, the success of these methods depends on a large amount of manually
labeled masks, which are expensive and time-consuming to be collected.
In this work, a novel Consistent PerceptionGenerative Adversarial Network (CPGAN) is proposed for semi-supervised stroke lesion
segmentation. The proposed CPGAN can reduce the reliance on fully labeled samples.
Specifically, A similarity connection module (SCM) is designed to capture the
information of multi-scale features. The proposed SCM can selectively aggregate the
features at each position by a weighted sum.
Moreover, a consistent perception strategy is introduced into the proposed model to
enhance the effect of brain stroke lesion prediction for the unlabeled data.
Furthermore, an assistant network is constructed to encourage the discriminator to
learn meaningful feature representations which are often forgotten during training stage.
The assistant network and the discriminator are employed to jointly decide whether the
segmentation results are real or fake.
The CPGAN was evaluated on the Anatomical Tracings of Lesions After Stroke (ATLAS).
The experimental results demonstrate that the proposed network achieves superior
segmentation performance.
In semi-supervised segmentation task, the proposed CPGAN using only two-fifths of
labeled samples outperforms some approaches using full labeled samples.
\end{abstract}


\section{Introduction}
\IEEEPARstart{S}{troke} is a leading cause of dementia and depression worldwide\cite{1_johnson2016stroke}.
Over two-thirds of stroke survivors experience long-term disabilities that impair
their participation in daily activities\cite{2_feigin2014global,3_kwakkel2003probability}.
Stroke lesion segmentation is the first and essential step of lesion recognition and
decision. Accurate identification and segmentation would improve the ability of
physicians to correctly diagnose patients.
Currently, the lesions are generally segmented manually by professional radiologists on
MR images slice-by-slice, which is time-consuming and relies heavily on subjective
perceptions\cite{4_liew2018large}.
Therefore, automatic methods for brain stroke lesion segmentation are in urgent demand
in the clinical practice. Nevertheless, there are great challenges with this task.
On the one hand, the scale, shape, size, and location of lesions limit the accuracy of
automatic segmentation. On the other hand, some lesions have fuzzy boundaries,
confusing the confidential partition between stroke and non-stroke regions.

With the development of machine learning in medical image
analysis\cite{54_wang2018automatic,55_wang2018skeletal,57_wang2018classification,wang2018bone,wang2020ensemble,yu2020multi},
automatic feature learning algorithms have emerged as feasible approaches for stroke
lesion segmentation.
Zhang et al.\cite{5_zhang2018automatic} proposed a 3D fully convolutional and
densely connected convolutional network (3D FC-DenseNet) for the accurate
automatic segmentation of acute ischemic stroke.
Bjoern at el.\cite{6_menze2015generative} designed a new generative probabilistic
model for channel-specific tumor segmentation in multi-dimensional images.
Hao et al.\cite{8_yang2019clci} designed a cross-level fusion and context
inference network(CLCI-Net) for the chronic stroke lesion segmentation from
T1-weighted MR images. Qi et al.\cite{7_qi2019x}
presented an end-to-end model named X-net for brain stroke lesion segmentation,
this approach achieved good performance on ATLAS.

Although many automatic segmentation methods have been presented, they are essentially
supervised learning methods.
Training a robust model requires a large number of manually labeled masks.
Due to the high cost for data labeling and patient privacy, it is difficult to collect
sufficient samples for training of the model in medical image analysis.
How to train an effective model using limited labeled data becomes an open and
interesting problem.

The recent success of Generative Adversarial Networks(GANs)\cite{47_goodfellow2014generative,001,002}
and application of variational inference\cite{56_mo2009variational,63,64,65} facilitates effective
unsupervised learning in numerous tasks.
The main reason is that GAN can automatically learn image characteristics in an
unsupervised manner.
Zhu et al.\cite{48_zhu2018adversarial} designed an end-to-end adversarial FCN-CRF
network for mammographic mass segmentation.
Zhao et al. \cite{49_zhao2018craniomaxillofacial} proposed a cascaded generative
adversarial network with deep-supervision discriminator (Deep-supGAN) for automatic
bony structures segmentation.
Lei et al. \cite{53_lei2020skin} adopted a effective GAN model for skin lesion segmentation
from dermoscopy images.
GAN has also been applied in the semi-supervised learning.
For instance, Zhang et al.\cite{52_zhang2017semi} proposed a novel semi-supervised method
to check the coverage of LV from CMR images by using generative adversarial networks.
Madani et al.\cite{51_madani2018semi} utilized a semi-supervised architecture of GANs to address
both problems of labeled data scarcity and data domain overfitting.
These studies have shown significant results by using both labeled data and
arbitrary amounts of unlabeled data.
However, the previous works focus mainly on the design of the generator and the use of
fake samples, but fail to take full advantage of the discriminator and the data itself.

Motivated by this, we propose a novel method named Consistent Perception GAN(CPGAN) for
semi-supervised segmentation task.
A similarity connection module is designed in the segmentation network to capture the
long-range contextual information, which contributes to the segmentation of lesions with
different shapes and scales. This module can aggregate the information of multi-scale
features and capture the spatial interdependencies of features.
The assistant network is proposed to improve the performance of discriminator using
meaningful feature representations.
More importantly, A consistent perception strategy is developed in adversarial training.
The rotation loss is adopted to encourage the segmentation network to make a consistent
prediction of the input, which contains rotated and original images.
A semi-supervised loss is designed according to the classification results of the
discriminator and the assistant network.
This loss can minimize the segmentation results between the labeled and unlabeled images.

In summary, the main contributions of this work can be listed as follows:
\begin{enumerate}
\item A non-local operation, SCM is designed to capture context information from
multi-scale features.
The proposed SCM can selectively aggregate the features at each position and enhance
the discriminant ability of the lesion areas.
\item The assistant network is employed to encourage the discriminator to learn
meaningful feature representations.
The assistant network and discriminator work together and this structure can improve
the performance of segmentation.
\item A consistent perception strategy is proposed to improve the recognition of the
unlabeled data. It makes full use of self-supervised information of the input and
encourages the segmentation network to predict consistent results. Our method using only
two-fifths of labeled samples outperforms some approaches using full labeled samples.
\end{enumerate}

The rest of the paper is organized as follows. Section II discussed the relevant
work about the architecture of the proposed network.
Section III discussed the architecture and distinctive characteristics of CPGAN.
Section IV summarized the results of extensive experiments including an ablation study
for the similarity connection module and the assistant network.
The quantitative and qualitative evaluations show that the proposed CPGAN achieves
better performance of stroke lesion segmentation and has good performance of
semi-supervised segmentation.
Finally,  Section V summarized the paper, and discusses some future research directions.

\begin{figure*}[htp]
  \centerline{\includegraphics[width=\textwidth]{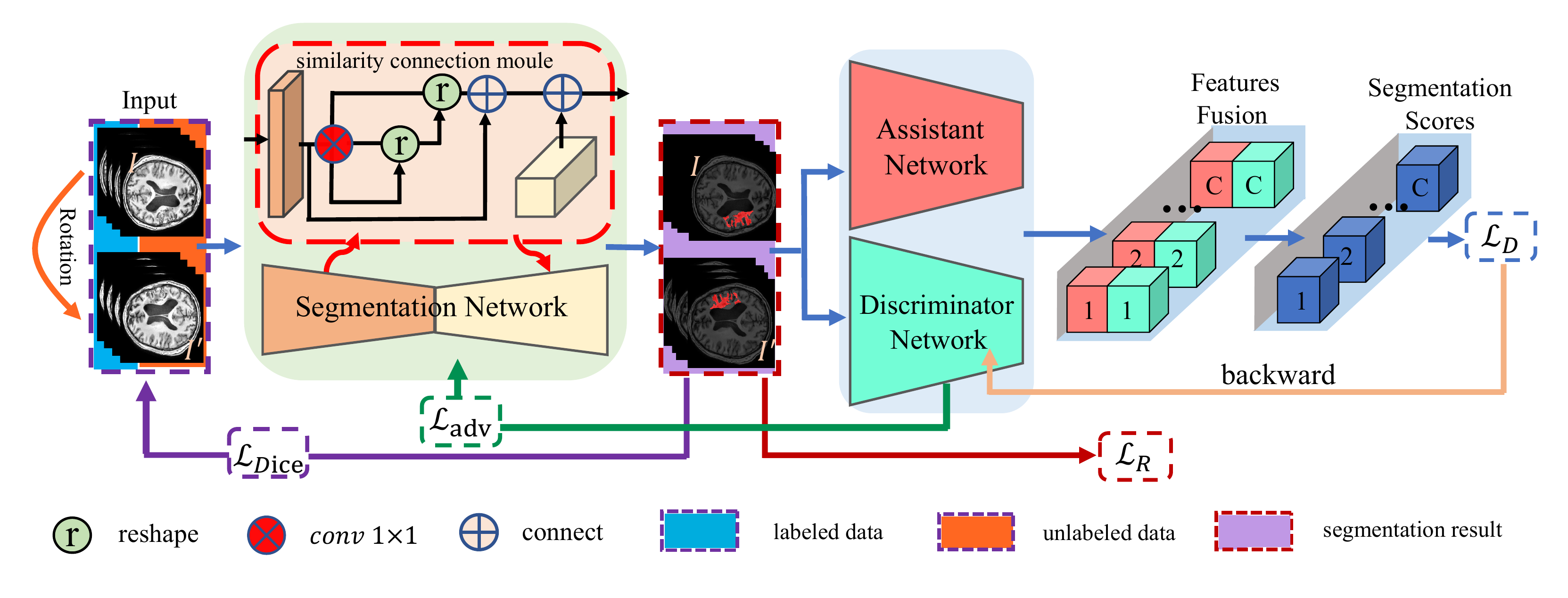}}
  \caption{The illustration of the architecture of Consistent Perception Generative Adversarial
  Network for brain stroke lesion segmentation. The assistant network is pre-trained and shares
  the same architecture as the discriminator. The segmentation network $\mathrm{S}$ is trained
  to predict the stroke lesion areas of original images and rotated images, while
  discriminator $\mathrm{D}$ is learned to distinguish between ground
  truth and original/rotated results of prediction with the help of the
  assistant network.}
\label{fig1}
\end{figure*}

\begin{figure*}[htp]
  \centerline{\includegraphics[width=\textwidth]{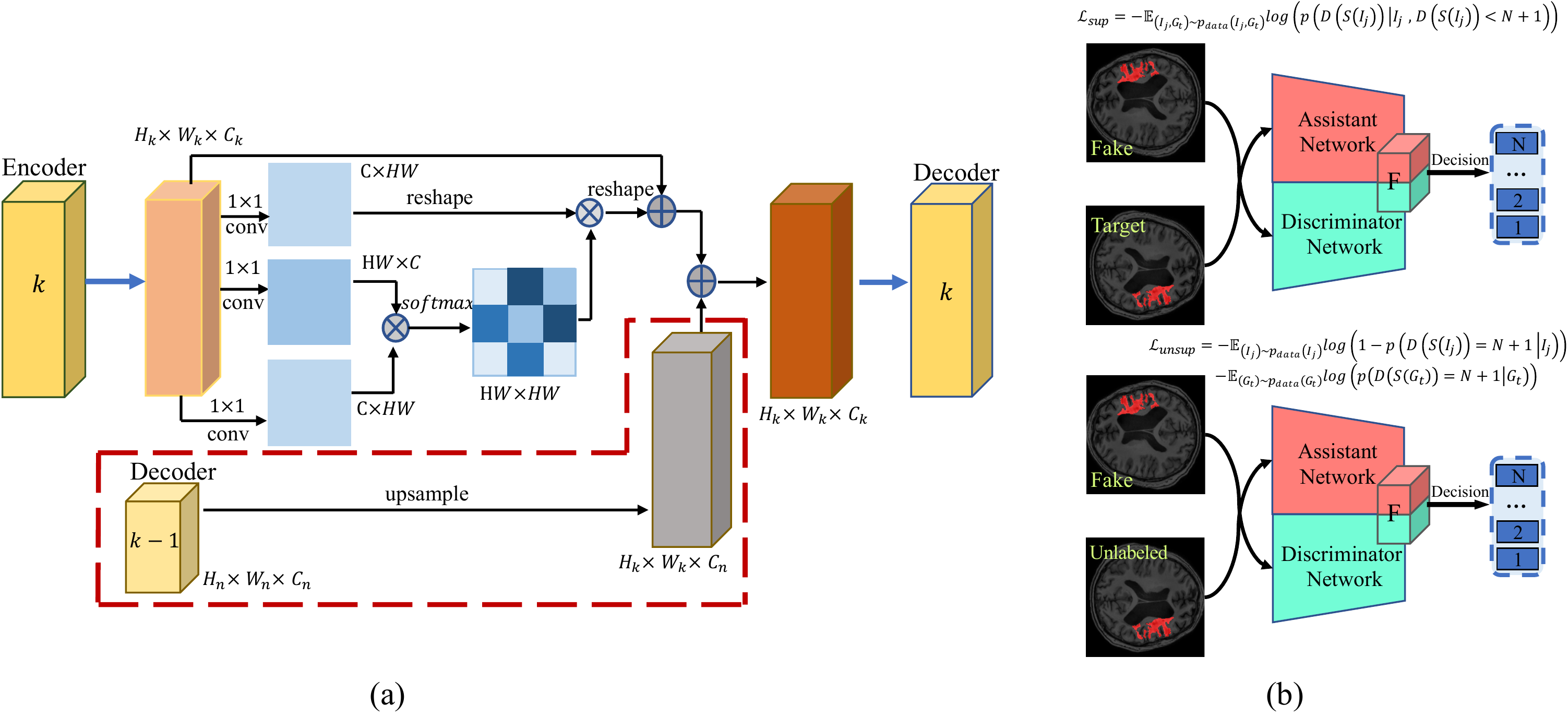}}
  \caption{(a)The detailed architecture of the proposed similarity connection module.
  The blue arrow indicates the process of skip connection. The red box indicates the
  process of the aggregating multi-scale feature map. (b)The discriminator and the assistant network work together.
  The feature map $\mathrm{F}$ is obtained by connecting two feature vectors, one from the last convolution layer of the assistant network and the other from the discriminator.
  Blue solid arrows represents forward and backward propagation, respectively. Representative features caused
  by the assistant network would be applied to the discriminator and affect
  discriminator update.
  }
\label{fig2s}
\end{figure*}

\section{Related Work}
\label{Related Work}

\subsection{U-net based methods}
Encoder-decoder architectures based segmentation methods have been widely used in image
segmentation task, such as U-net \cite{10_ronneberger2015u},
H-DenseU-net \cite{11_li2018h},U-net++ \cite{12_zhou2018unet++}.
It has becomes a popular neural network architecture for biomedical image segmentation
tasks \cite{13_dunnhofer2020siam,14_hiasa2019automated,15_man2019deep,16_zhu2019boundary}.
Huang et al. \cite{17_huang2017densely} introduced a dense convolutional network (DenseNet)
with dense blocks, which created short paths from the early layers to the latter layers.
Baur et al. \cite{18_baur2017semi} proposed a semi-supervised learning framework for
domain adaptation with embedding technique on challenging task of Multiple Sclerosis
lesion segmentation.
Sedai et al. \cite{19_sedai2017semi} introduced a generative variational autoencoder
that was trained using a limited number of labeled samples had a good performance of
optic cup segmentation. Huang et al. \cite{20_huang2020unet} proposed a full-scale
connected model Unet3 +, which has better segmentation performance for organs of
different sizes in various medical images.
However, existing models rely on an encoder-decoder architecture with stacked local
operators to aggregate long-range information gradually.
Those methods are easy to cause the loss of spatial information\cite{21_gu2019net}.

To address this issues, Ozan et al. \cite{22_oktay2018attention} adopted a novel
attention gate (AG) model for medical imaging that automatically learns to focus on
target structures of varying shapes and sizes.
Nabila et al. \cite{23_abraham2019novel} combined attention gated U-Net with a novel
variant of the focal Tversky loss function to address the problem of data imbalance in
medical image segmentation.
Wang et al. \cite{24_wang2020non} introduced the non-local U-Nets equipped with flexible
global aggregation blocks, this method outperforms previous models significantly with
fewer parameters and faster computation on the 3D multimodality isointense infant
brain MR image segmentation task.
Fu et al. \cite{25_fu2019dual} proposed a dual attention network to integrate
local features with global dependencies, their appended the position attention module
and channel attention module on the top of FCN and achieved good performance on three
challenging scene segmentation dataset.
Unfortunately, those skip connections demand the fusion of the same-scale encoder and
decoder feature maps\cite{26_hasan2019u}, and those methods is insensitive to the
different sizes and locations of lesions.
Inspired by these previous studies, we proposed a segmentation module using a similarity
connection module to enhance the ability of the representation in our CPGAN.

\subsection{GAN based methods}
Generative Adversarial Networks(GANs), which can be derived via variational inference
consist of two components parts: a generative network that generates pseudo data,
and a discriminative network that differentiates between fake and real data.
Recently, GAN has gained a lot of attention in the medical image computation field
\cite{57_wang2018classification,58_wang2020diabetic,59_yu2021tensorizing,60_hu2020brain,61_you2020fine,62_hu2020medical}
due to its capability of data generation. These properties have been adopted in many
segmentation methods \cite{27_wolterink2017generative,28_dar2019image,29_yang2018low,30_yu2019ea,31_Sharma2019Missing,32_Michael2019Generative,33_chen2019ganpop}.
For example, Nie et al. \cite{34_nie2018strainet} proposed a spatially-varying
stochastic residual adversarial network (STRAINet) to delineate pelvic organs from MRI
in an end-to-end fashion, this model achieved significant improvement in pelvic organ
segmentation. Chen et al. \cite{35_chen2019one} constructed a one-shot generative
adversarial learning framework to make full use of both paired and unpaired MRI-CT data
for MRI segmentation of craniomaxillofacial(CMF) bony structures.
Micheal et al. \cite{36_gadermayr2019generative} proposed a fully unsupervised
segmentation approach exploiting image-to-image translation to convert from the image
to the label domain.
Xue et al. \cite{37_xue2018segan} proposed a novel end-to-end GAN, called SegAN,
with a new multiscale loss for the task of brain tumor segmentation.

This adversarial training scheme and framework are also used for semi-supervised
segmentation task. Chen et al.\cite{38_chen2019multi} proposed a semi-supervised method
called MASSL that combines a segmentation task and a reconstruction task through an
attention mechanism in a multi-task learning network.
Zheng et al. \cite{39_zheng2019semi} used adversarial learning with deep atlas prior
to do semi-supervised segmentation of the liver in CT images.
This method utilized unannotated data effectively and achieved good performance on ISBI
LiTS 2017. However, the created pseudo labels usually do not have the same quality as
the ground truth for the target segmentation objective, it would limit the performance
of the semi-supervised segmentation models.
Based on GAN structure, we introduce an assistant network and consistent perception
strategy to improve the semi-supervised performance.

\section{Consistent Perception Generative Adversarial Network}

The proposed semi-supervised learning method is shown in Fig. \ref{fig1}.
CPGAN consists of three neural networks: segmentation network, discriminator network
and assistant network. We adopt U-net architecture and equip skip connections with
similarity connection module in the segmentation network to improve segmenting
performance. Rotated images and original images input to the segmentation network to
predict segmentation results.
The network equivariant property\cite{40_worrall2017harmonic} is utilized to obtain
rotation loss and semi-supervised loss from labeled and unlabeled images.

\subsection{The architecture of Segmentation Network}

In the generator, U-net is applied to extract features.
This architecture is composed of a down-sampling encoder and an up-sampling decoder.
Skip connections are adopted to aggregate the same-scale feature map and capture local
and global contextual information.
However, a common limitation of the U-Net and its variations is that the consecutive
pooling operations or convolution striding reduce the feature resolution to learn
increasingly abstract feature representations.
To address this challenge, a similarity connection module is proposed to extract a wide
range of sensitive position information and multi-scale features.
We append this non-local operation on the skip connections and sum the local information
of varying scales at a decoder node.

The architecture of similarity connection module is shown in Fig.\ref{fig2s}(a).
A local feature map $\mathrm{E} \in R^{H} k^{\times W}_{k} \times C_{k}$
is fed into a convolution layer which generates two new feature maps
$\{B, F\} \in R^{H} k^{\times W_{k} \times C_{k}}$. Those feature maps are reshaped to
$R^{\mathrm{I}_{k} \times C_{k}}$, where $I=H_{k} \times W_{k}$ and $\mathrm{k}$,
is the number of layers in the Encoder. To compute the relationship of
$\left(B_{i}, F_{j}\right)$, for each pair of position, a matrix multiplication
$q\left(B_{i}, F_{j}\right), i \in H_{k}, j \in W_{k}$ is calculated between
$\mathrm{B}$ and $\mathrm{F}$, $q\left(B_{i}, F_{j}\right)$ consists of a softmax layer
and measures the $i^{t h}$ position's impact on $j^{t h}$ position:

\begin{equation}
  q\left(B_{i}, F_{j}\right)=\frac{\exp \left(B_{i}, F_{j}\right)}{\sum_{j}^{N} \exp \left(B_{i}, F_{j}\right)}.
\end{equation}

Meanwhile, $\mathrm{E}$ is fed into a $1 \times 1$ convolution layer to
generator a new feature map $\mathrm{P} \in R^{H} k^{\times W_{k} \times C_{k}}$
and reshaped to $R^{\mathrm{I}_{k} \times C_{k}}$. Then, a matrix multiplication
is calculated between $q\left(B_{i}, F_{j}\right)$ and $\mathrm{P}$.
The result  $\mathrm{E}^{*}$ is reshaped to $R^{H_{k} \times W_{k} \times C_{k}}$.
An upsample operation is performed on $D \in R^{H_{n} \times W_{n} \times C_{n}}$ to
get a new feature map $D^{*} \in R^{H} k^{\times W_{k} \times C_{n}}$, where $n=k-1$.
Finally, we sum $\mathrm{E}^{*}$ and $\mathrm{D}^{*}$ to obtain the output of similarity
connection module:

\begin{equation}
S_{i}=\sum_{j=1}^{N} q\left(B_{i}, F_{j}\right) \cdot P+D^{*}=\sum_{j=1}^{N} \mathrm{E}^{*}+D^{*}.
\end{equation}

Therefore, the result of similarity connection module $S_{i}$ can capture long-range
contextual information according to non-local attention map and aggregate multi-scale
feature to decrease the loss of spatial information.

\subsection{The architecture of Assistant Network}

The original value function for GAN training is:

\begin{equation}
  \begin{aligned}
      V(G, D)=& \mathbb{E}_{\boldsymbol{x} \sim P_{\text {data }}(\boldsymbol{x})}\left[\log P_{D}(S=1 \mid \boldsymbol{x})\right] \\
      &+\mathbb{E}_{\boldsymbol{x} \sim P_{G}(\boldsymbol{x})}\left[\log \left(1-P_{D}(S=0 \mid \boldsymbol{x})\right)\right],
  \end{aligned}
\end{equation}

Where $P_{\text {data }}$ is the true data distribution,
$P_{G}$ is the generator's distribution, and $P_{D}$ is the discriminator's distribution.
Training is typically performed via alternating stochastic gradient descent.
Therefore, at iteration k during training, the discriminator classifies samples as
coming from $P_{\text {data }}$ or $P_{G}$ .
As the parameters of G change, the distribution $P_{G}$ changes, which implies a
non-stationary online learning problem for the
discriminator\cite{41_bang2018improved,44_chen2019self-supervised}.
To address this challenge, we propose an assistant network to prevent this forgetting
of the classes in the discriminator representations.

As shown in Fig.\ref{fig2s}(b), the assistant network is pre-trained and shares the
same architecture as the discriminator.
Assistant network-parameters are fixed during the training, the predicted segmentation
results $S\left(I_{j}\right)$ input the assistant and discriminator network.
We concatenate two feature maps, one from the last layer of the assistant network and
the other from the discriminator.
We can derive the gradient toward the discriminator by calculating the partial
derivative of loss term $W_{D}$:

\begin{equation}
  \begin{aligned}
      \nabla W_{D} &=\frac{\partial D\left(S\left(I_{j}\right)\right)}{\partial h_{2}} \cdot \frac{\partial h_{2}}{\partial W_{D}} \\
      &=-\frac{1}{Y} \cdot Y(1-Y) \cdot W_{2} \cdot \frac{\partial h_{2}}{\partial W_{D}} \\
      &=(Y-1) \cdot W_{2} \cdot u\left(W_{D}\right),
  \end{aligned}
\end{equation}
Where $D\left(S\left(I_{j}\right)\right)=-\log Y, Y=\sigma\left(h_{1} w_{1}+h_{2} w_{2}\right)$ and $Y$ is
softmax function. $w_{1}$ and $w_{1}$ are network parameters. $h_{1}$ and $h_{2}$ represent assistant and
discriminative feature maps, respectively. $u\left(W_{D}\right)$ is gradient update formula of
discriminator's penultimate layer. Thus, $\nabla W_{D}$ depends on $h_{1}$. The representative
features affect the discriminator update.
Therefore, the generator is trained by considering both assistant and
discriminative features, because it should fool the discriminator by maximizing
$-\log \left(D\left(S\left(I_{j}\right)\right)\right)$. Representative information of the assistant
network would help the model to learn the information of stroke lesion faster and converge quickly.
\subsection{Training Strategy and Loss Functions}

To improve the generalization capability of the network, the transformation equivariance
has been proposed. Cohen and Welling \cite{42_cohen2016group} proposed group equivariant
neural network to improve the network generalization.
Dieleman et al. \cite{43_dieleman2016exploiting} designed four different equivariance
to preserve feature map transformations by rotating feature maps instead of filters.
Chen et al. \cite{44_chen2019self-supervised} presented an unsupervised generative model
that combines adversarial training with self-supervised learning by using auxiliary
rotation loss.
Inspired by these works, our consistent perception strategy targets to utilize the
unlabeled images better in semi-supervised learning.

In the consistent perception strategy of the proposed method, labeled and unlabeled
images are rotated.
Rotated images and original images are added up to the input. Segmentation is desired as
transformation equivariant. If the input images are rotated, the prediction of ground
truth masks should be rotated in the same way compared to original masks.
Rotation loss $\mathcal{L}_{R}$ is adopted to evaluate the equivariant representation of
segmentation network output on both labeled and unlabeled image, which is obtained by:

\begin{equation}
  \mathcal{L}_{R}=-\sum_{j \in N} E\left(R\left(I_{j}\right)-I_{j}\right)^{2}.
\end{equation}

The segmentation network equipped with similarity connection module is trained
using Dice loss $\mathcal{L}_{\text {Dice }}$ on labeled images only:
\begin{equation}
  \mathcal{L}_{\text {Dice }}=-\sum_{j \in N} \sum G_{t} \log \left(S\left(I_{j}\right)\right),
\end{equation}

Where $\mathcal{L}_{\text {Dice }}$ and $\mathcal{L}_{R}$ are cross-entropy loss and
mean square error. $G_{t}$ is the ground truth label and $I_{j}$ is input images,
$S(\cdot)$ and $R(\cdot)$ denote the segmentation network and rotation operation,
and $S\left(I_{j}\right)$ denotes the predicted results.

For unlabeled images, the rotation loss encourages the segmentation network to learn more self-supervised information. The adversarial loss brought by the discriminator provides a clever way of unlabeled samples into training. Some semi-supervised learning methods of GANs use discriminator for $N+1$ classification and treat generated data as $N+1$. Instead, we exploit a novel technique to discriminate prediction of segmentation network. Ground truth, rotated and original labeled data are judged as $ N+1 $, $ N $, $N-1$, respectively. In this paper, $ N=2 $.

Therefore, our loss function for training the discriminator $\mathcal{L}_{D}$ is:
\begin{equation}
  \mathcal{L}_{D}=\mathcal{L}_{\text {sup}}+\mathcal{L}_{\text {unsup}},
\end{equation}

\begin{equation}
  \begin{aligned}
  \mathcal{L}_{\text {sup}}&=-\mathbb{E}_{\left(I_{j}, G_{t}\right) \sim p_{\text {data}}\left(I_{j}, G_{t}\right)} \\
&\quad\log \left(p\left(D\left(S\left(I_{j}\right)\right) | I_{j}, D\left(S\left(I_{j}\right)\right)<N+1\right)\right),
  \end{aligned}
\end{equation}

\begin{equation}
  \begin{aligned}
  &\mathcal{L}_{\text {unsup}}=-\mathbb{E}_{\left(I_{j}\right) \sim p_{\text {data }}\left(I_{j}\right)}\log \left(1-p\left(D\left(S\left(I_{j}\right)\right)=N+1 | I_{j}\right)\right) \\
  &\qquad\qquad-\mathbb{E}_{\left(G_{t}\right) \sim p_{\text {data }}\left(G_{t}\right)}\log \left(p\left(D\left(S\left(G_{t}\right)\right)=N+1 | G_{t}\right)\right).
  \end{aligned}
\end{equation}

Meanwhile, we use the adversarial learning to improve the performance of segmentation.
With the loss $\mathcal{L}_{a d v}$ segmentation network is training to fool the
discriminator by maximizing the probability of the prediction masks.
This loss is generated from the distributions of ground truth masks:

\begin{equation}
  \mathcal{L}_{a d v}=-\mathbb{E}_{\left(I_{j}\right) \sim p_{d a t a}\left(I_{j}\right)} \log \left(p\left(D\left(S\left(I_{j}\right)\right)=N+1 | I_{j}\right)\right).
\end{equation}

In summary, the proposed rotation perception strategy training strategy encourages the
discriminator to learn useful image representation and detects the rotated transformation.
This semi-supervised training method promotes segmentation network to make the same
prediction on both labeled and unlabeled images.

\section{Experiments and Results}

\subsection{Data and Evaluation Metrics}
\subsubsection{Dataset}
The CPGAN is evaluated on an open dataset, Anatomical Tracings of Lesions After Stroke
(ATLAS), which contains 239 T1-weighted normalized  3D MR images with brain stroke
lesion manually labeled mask. We randomly selected 139 subjects for training, 40 for
validation and 60 for testing.
Each of objects is cropped to 189 slices which size is "233$\times$197" .
In order to keep the size of rotated images consistent, slices are expended
to "256$\times$256" . Noted that only in the training dataset, the input images are
rotated by $180^{\circ}$.

\subsubsection{Evaluation Metrics}

In this paper, we employ 4-fold cross-validation strategy and use a series of evaluation
metrics to measure the performance of our model, including Dice coefficient(Dic),
Jaccard index(Jac), Accuracy(Acc), Sensitivity(Sen) and Specificity(Spe).
The definition of them are:
\begin{equation}
  Dic=\frac{2 \cdot T P}{2 \cdot T P+F N+F P},
\end{equation}

\begin{equation}
  Jac=\frac{ T P}{ T P+F N+F P},
\end{equation}
\begin{equation}
  Acc=\frac{T P+T N}{T P+F P+T N+F N},
\end{equation}
\begin{equation}
  Sen=\frac{T P}{T P+F N},
\end{equation}

\begin{equation}
  Spe=\frac{T N}{T N+F P},
\end{equation}

Where $TN$,$TP$,$FN$ and $FP$ refer to the number of true negatives, true positives,
false negatives, false positives, respectively.

\subsubsection{Implementation}
Our implementation is based on Pytorch. 4 NVIDIA RTX 2080Ti with 11 GB memory are
used for each experiment with a batch size of 4. The proposed network was trained
with a fixed learning rate of 0.001. The strategy of reduce learning rate is adopted to reduce learning rate automatically and the Adam optimizer is used to minimize the loss function.

\subsection{Ablation Analysis of Similarity Connection Module}
We employ the similarity connection module(SCM) to extract a wide range of sensitive
position information and multi-scale features. Three methods are used to conduct a
comparative experiment. The similarity connection module is added into the architecture
of U-net, ResUnet, and CPGAN, respectively.
In Table \ref{tab1}, it can be observed that employing SCM gains better performance in
five evaluation metrics compared to the original method. Equipped with SCM,
Unet-SCM(U-net with similarity connection module) performs better with 0.076, 0.061,
0.089, 0.053 and 0.116 improvement on Dice, Jaccard index,
Accuracy, Sensitivity and Specificity, respectively. ResUnet-SCM(ResUnet with the
similarity connection module) performs better with 0.069, 0.126, 0.060, 0.088 and 0.074
improvement on Dice, Jaccard index, Accuracy, Sensitivity and Specificity, respectively.
Compared with Unet-SCM and ResUnet-SCM, similarity connection module is more effective
in CPGAN with 0.103, 0.160, 0.109, 0.131 and 0.149 improvement on Dice, Jaccard
index, Accuracy, Sensitivity and Specificity, respectively.
All results show that the proposed module performs very well in Sensitivity and
Specificity. It is worthwhile getting a higher score in these two evaluation metrics for
brain stroke segmentation tasks, because we need to make sure that all the strokes
can be detected and prevent non-diseased areas are misdiagnosed as brain stroke.

\begin{table}
  \renewcommand\arraystretch{1.5}
  \begin{center}
  \setlength{\abovecaptionskip}{0pt}%
  \setlength{\belowcaptionskip}{10pt}%
  \caption{Ablation study on ATLAS dataset for Similarity Connection Module}
  \label{tab1}
  \normalsize
  \setlength{\tabcolsep}{2mm}{
  \begin{tabular}{ccccccc}
  \toprule
  Method & SCM & Dic   & Jac   & Acc   & Sen   & Spe   \\ \midrule
  \multirow{2}*{U-net}  &     & 0.468 & 0.374 & 0.542 & 0.440  & 0.573 \\
         & \checkmark   & 0.544 & 0.435 & 0.631 & 0.493 & 0.689  \\ \hline
  \multirow{2}*{ResUnet}  &     & 0.470 & 0.351 & 0.556 & 0.427  & 0.602 \\
         & \checkmark   & 0.539 & 0.477 & 0.616 & 0.515 & 0.676  \\ \hline
  \multirow{2}*{CPGAN}  &     & 0.514 & 0.421 & 0.529 & 0.425 & 0.556 \\
         & \checkmark   & 0.617 & 0.581 & 0.638 & 0.556 & 0.705 \\
  \bottomrule
  \end{tabular}}
  \end{center}
  \label{table1}
\end{table}

\begin{figure*}[htp]
  \centerline{\includegraphics[width=\textwidth]{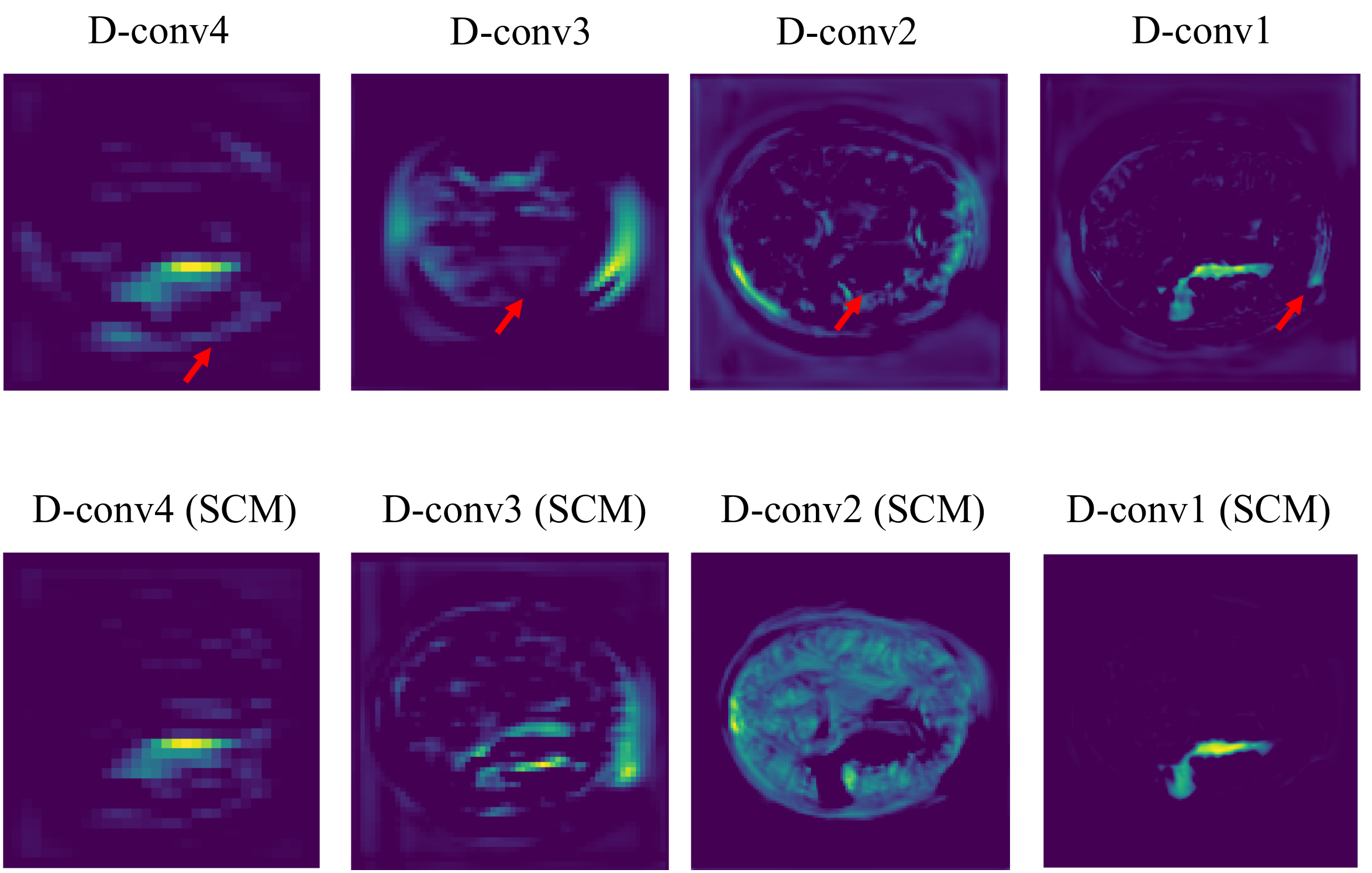}}
  \caption{Intermediate feature visualization of proposed CPGAN.
  These feature maps illustrate that the network can capture effective information
  of stroke lesion.}
\label{fig3}
\end{figure*}

\begin{figure*}[htp]
  \centerline{\includegraphics[width=\textwidth]{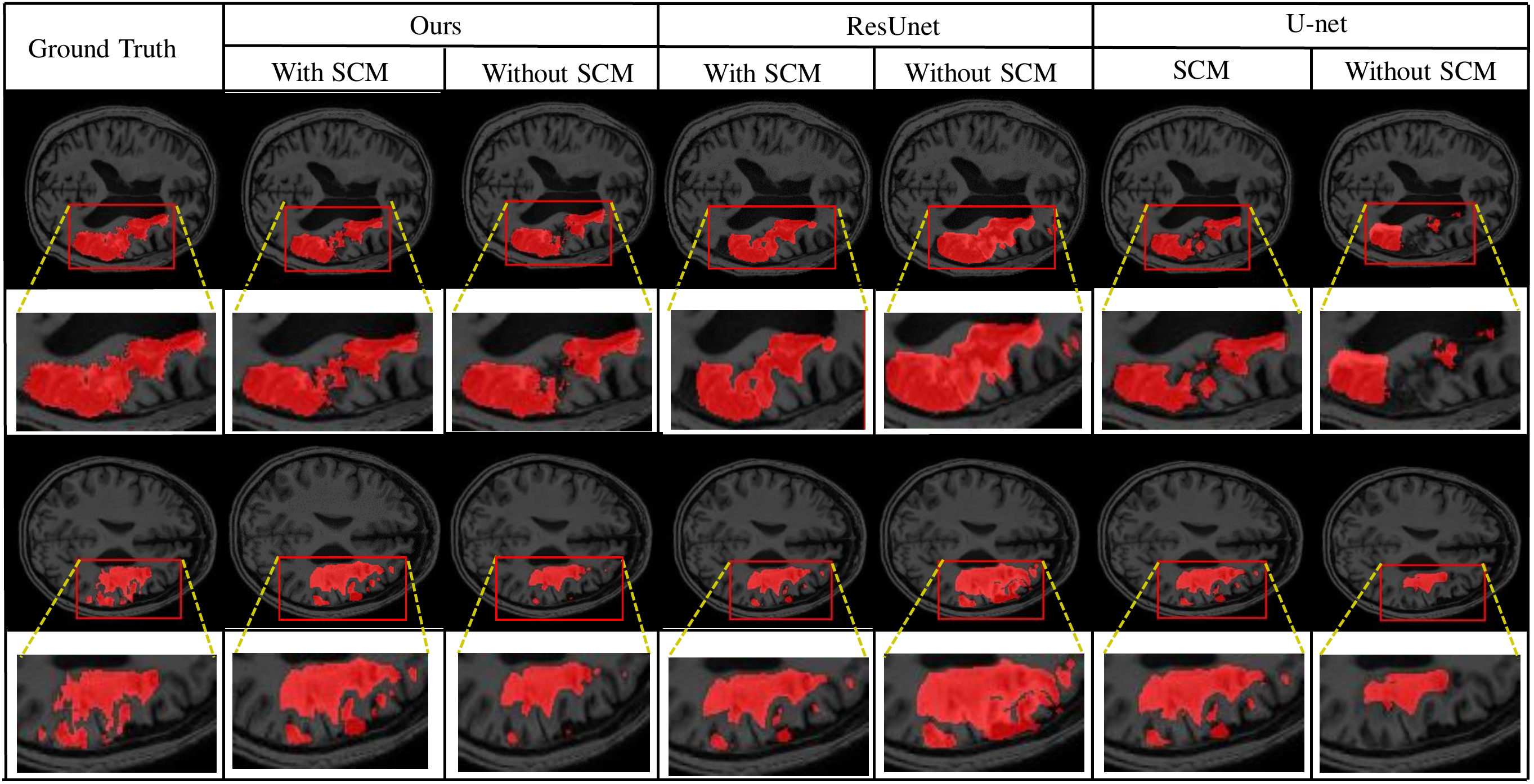}}
  \caption{Results of ablation study on the proposed similarity connection module based on ATLAS dataset, }
\label{4_result}
\end{figure*}

To understand the advantages of the similarity connection module better,
we further present the visualization results of the features from the Decoder
part of CPGAN unequipped with SCM in Fig.\ref{fig3}.
From this figure, the original Decoder part has two problems.
For one thing, model learns redundant information in non-lesion areas.
For another, the valid information is not captured in lesion areas.
Compared with the CPGAN. SCM has ability of capturing long-range contextual information
of the stroke lesion areas and reducing the learning of the redundant information.

The segmentation results of different methods are shown in Fig.\ref{4_result}.
More details are captured with our proposed similarity connection module.
It is demonstrated that the proposed module can help model achieve better performance
of segmentation consistently, and some of the interdependencies might have already been
captured with our proposed SCM.

\begin{figure}[htp]
  \centerline{\includegraphics[width=0.5\textwidth]{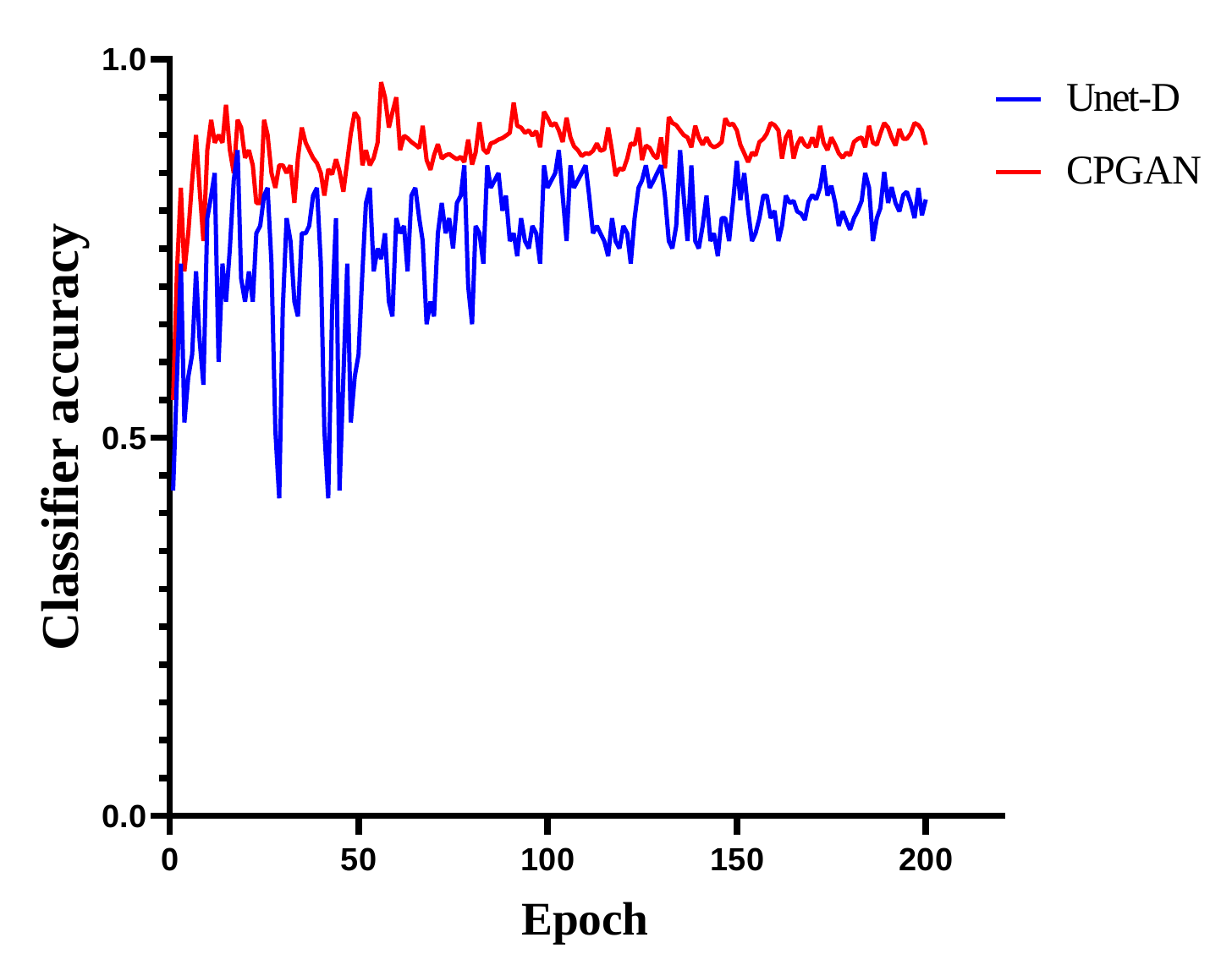}}
  \caption{The red line represents the change of discriminator's
  classification accuracy after the addition of assistant network,
  while the blue represents the change of the original model.}
\label{1}
\end{figure}

\begin{figure}[htp]
  \centerline{\includegraphics[width=0.5\textwidth]{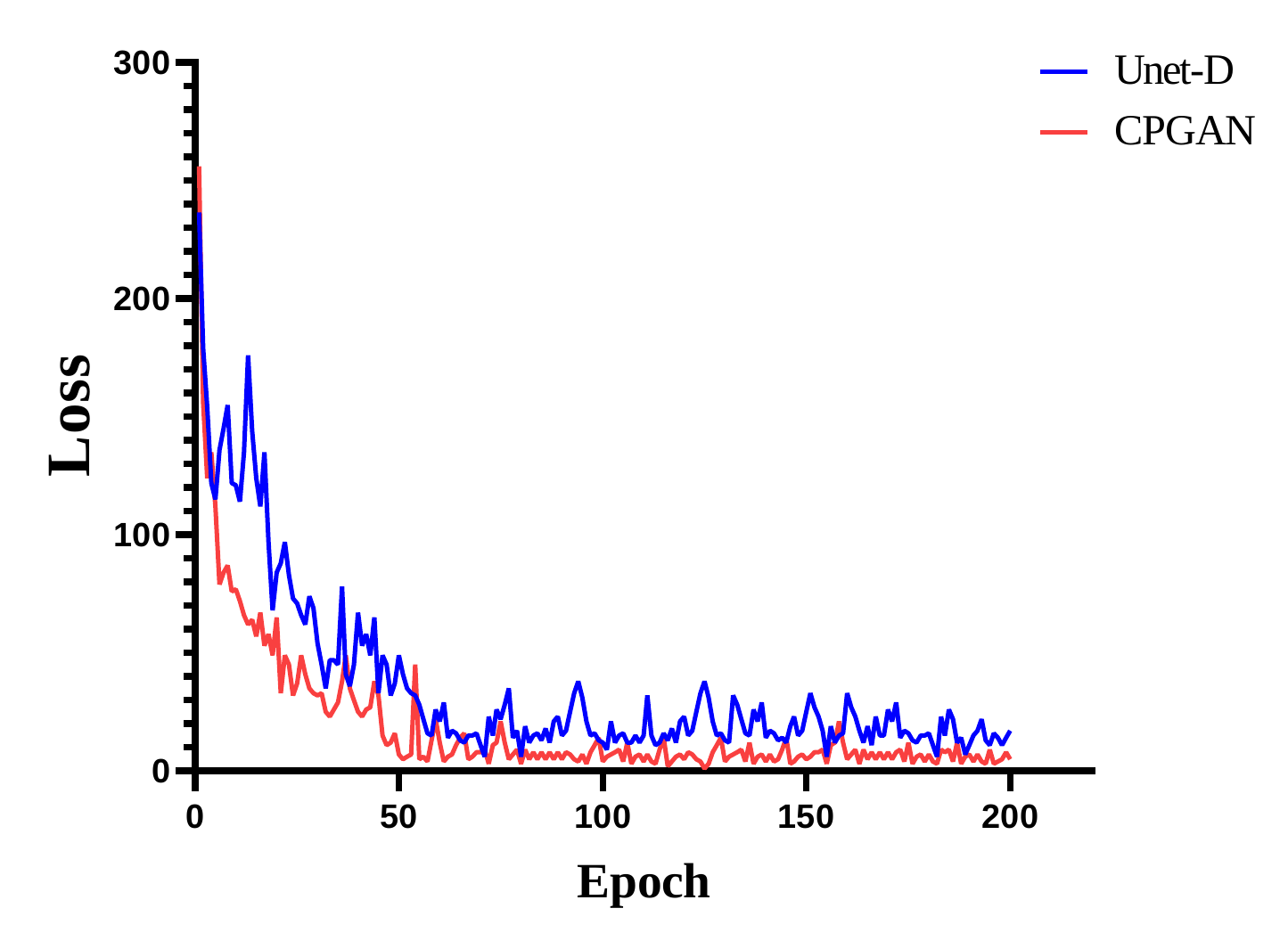}}
  \caption{The red line represents the change of the loss of segmentation network
  after adding assistant network, while the blue represents the change of the original
  model.}
\label{2}
\end{figure}

\subsection{Analysis of Assistant Network}

A set of experiments are conducted to validate the effectiveness of the proposed
assistant network. To verify the effectiveness of this network, we conduct experiments
with two models: CPGAN, Unet-D unequipped with the assistant network, the rest of the
Unet-D's structure is the same as CPGAN. It could be clearly observed from Fig.\ref{1}
and Fig.\ref{2} that employing assistant network enhances the performance of the model.
Fig.\ref{1} shows the change of the classification accuracy of the discriminator.
The Unet-D is unstable from 0 to 50 epochs. The classification accuracy drops
substantially every 5 epochs.
After 150 epochs, the classification accuracy is beginning to stabilize.
This demonstrates that the discriminator does not retain useful information in this
non-stationary environment. Some representations of lesion areas are forgotten during
training and this forgetting correlates with training instability.
After adding assistant network, we observe that proposed method can mitigate this problem.
Compared with Unet-D, the classification accuracy of CPGAN is stable after 20 epochs and
the performance improves by an average of 10 percent. Representative information of
assistant network improves the discriminator to learn meaningful feature representations.
It can be verified that assistant
network improves performance of CPGAN. As shown in Fig.\ref{2}, we can observe that the loss of
segmentation network changes. CPGAN decreases faster than Unet-D. After 50 epochs, the loss of
CPGAN tends to be stable, while Unet-D requires 70 epochs. This experiment indicates that adopting
the proposed assistant network can help discriminator mitigate the problem of discriminator forgetting.
It can be inferred that our proposed method converges quickly and achieves better performance.

\subsection{Rotation Perception strategy Strategy and Semi-supervised segmentation}

\begin{table*}
\renewcommand\arraystretch{1.5}
\begin{center}
\caption{Comparison of brain stroke segmentation results on ATLAS dataset}
\label{tab2}
\normalsize
\setlength{\tabcolsep}{4mm}{
\begin{tabular}{cccccc}
\toprule
Method      & Dic   & Jac   & Acc   & Sen   & Spe   \\\midrule
DeepLab V3+\cite{45_chen2018encoder} & 0.487 & 0.392 & 0.571 & 0.523 & 0.585 \\
Dense U-net\cite{10_ronneberger2015u} & 0.538 & 0.466 & 0.628 & \textbf{0.562} & 0.657 \\
U-net\cite{10_ronneberger2015u}       & 0.468 & 0.374 & 0.542 & 0.440  & 0.573 \\
DCGAN\cite{46_li2017brain}      & 0.439 & 0.388 & 0.529 & 0.425 & 0.556 \\
X-Net\cite{7_qi2019x}      & 0.572 & 0.457 & \textbf{0.646} & 0.493 & 0.679 \\
CPGAN       & \textbf{0.617} & \textbf{0.581} & 0.638 & 0.556 & \textbf{0.705}\\
\bottomrule
\end{tabular}}
\end{center}
\end{table*}

\subsubsection{Comparison to state-of-the-art methods}

We compare CPGAN with different state-of-the-art segmentation methods, including U-net,
DenseU-net(2D), DeepLab V3+\cite{45_chen2018encoder}, DCGAN\cite{46_li2017brain}
and X-net\cite{7_qi2019x}.
We briefly introduce these models here and the details can be found in the references.
DCGAN applies the convolutional operators to replace the pooling operators, strided
convolutions for the discriminator and fractional strided convolutions for the generator.
X-net adds a feature similarity module and a X-block to the U-net based architecture,
it is the top three methods on ATALS dataset leaderboard. From the results listed in
Table \ref{tab1}, it can be clearly observed that the proposed model scores are the
highest on the main indicators.
Besides, Table \ref{tab2} shows that our CPGAN performs better than other methods and
our segmentation method makes significant improvement than \cite{7_qi2019x}.
Although our method is not as good as other methods in Accuarcy and sensitivity,
it delivers promising performance on other evaluation metrics.
Compared with X-net, our method performs better with 0.045, 0.124 and 0.026
improvement on Dice, Jaccard and specificity, respectively.

\begin{figure}[htp]
  \centerline{\includegraphics[width=0.5\textwidth]{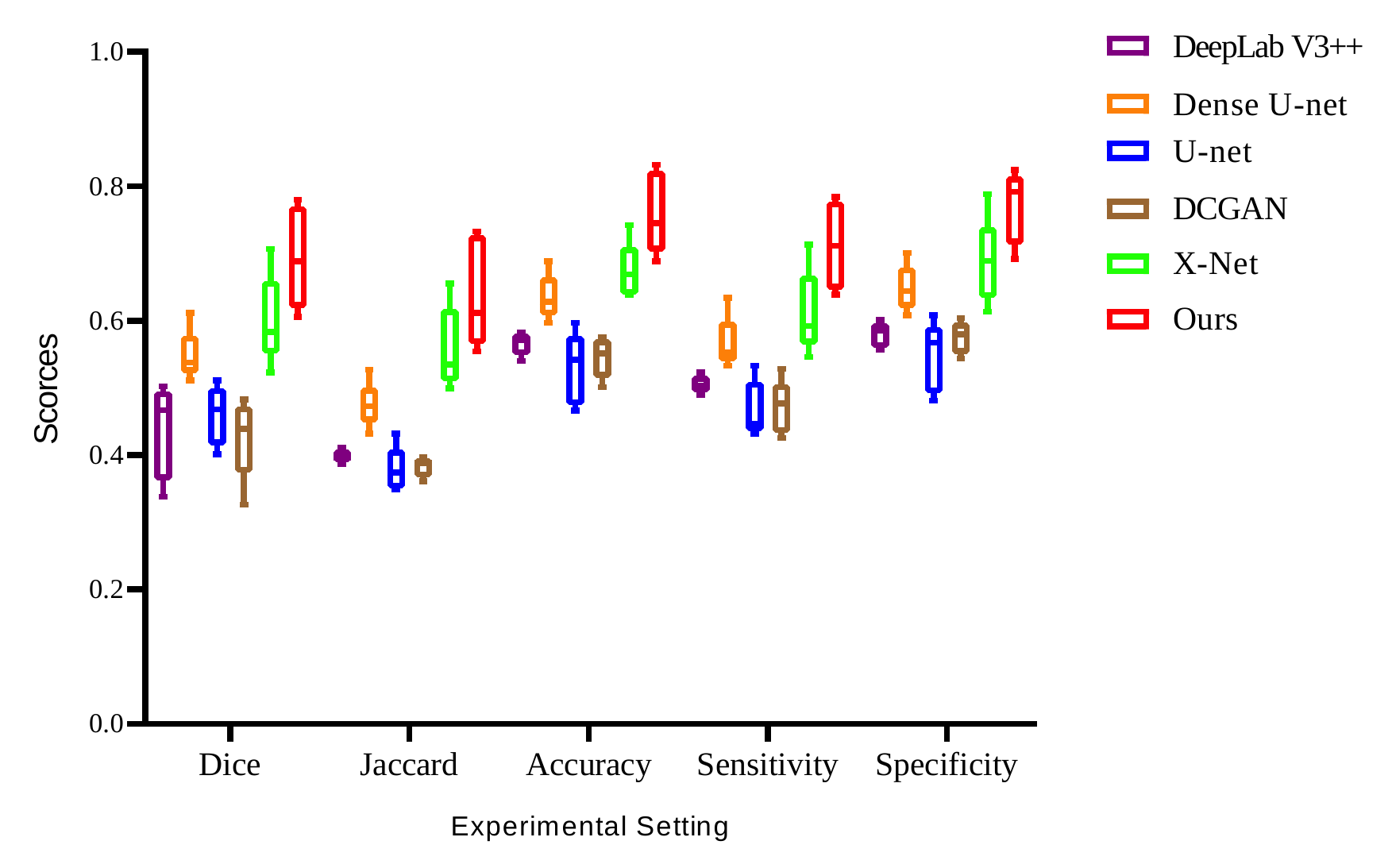}}
  \caption{Box-plots of Dice, Jaccard index, Accuracy, Sensitivity and Specificity of different models.}
\label{3}
\end{figure}

\begin{figure*}[htp]
  \centerline{\includegraphics[width=0.9\textwidth]{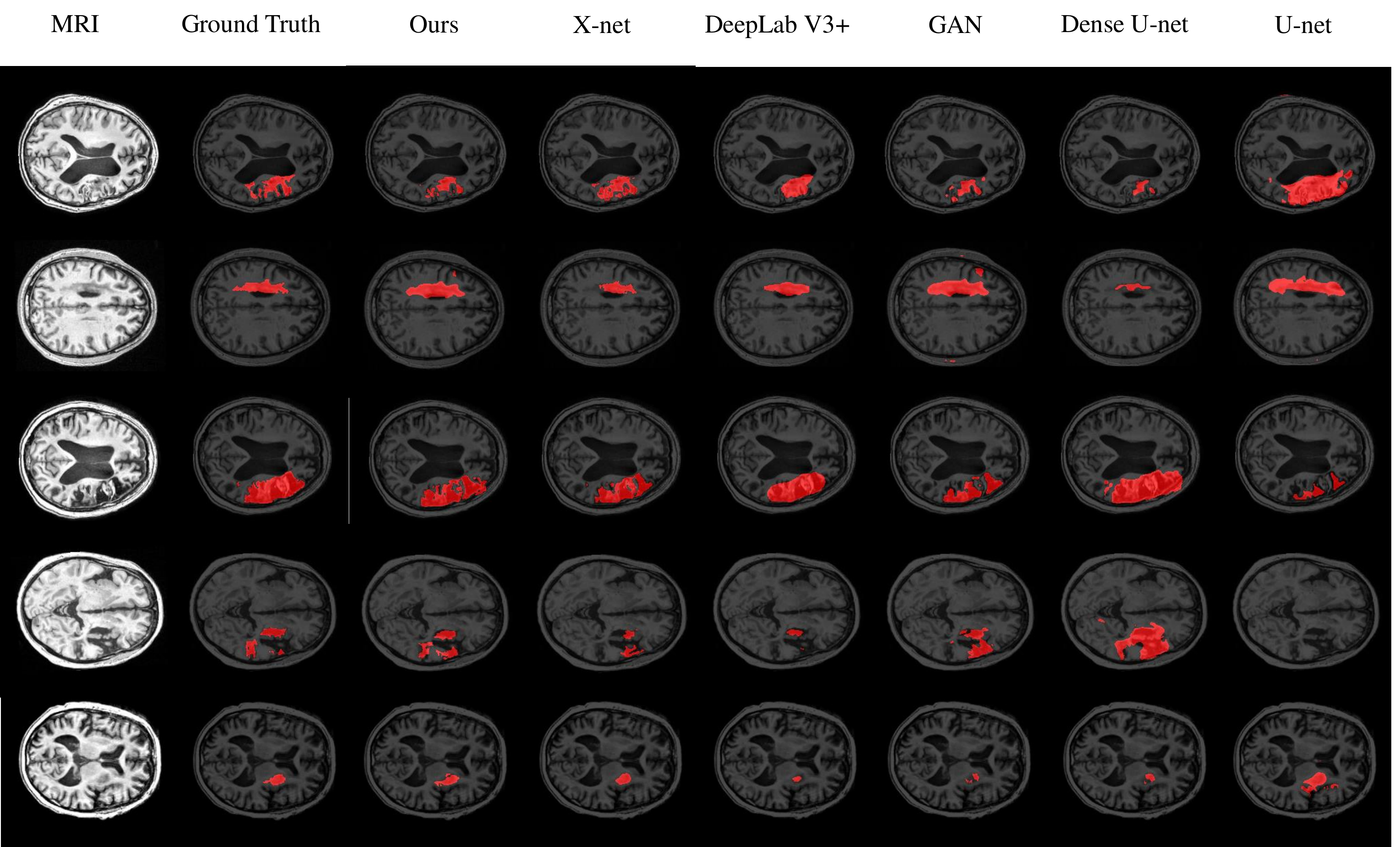}}
  \caption{Comparison between the proposed CPGAN and other state-of-the-art methods.}
\label{6_vs}
\end{figure*}

\begin{figure*}[htp]
  \centerline{\includegraphics[width=0.9\textwidth]{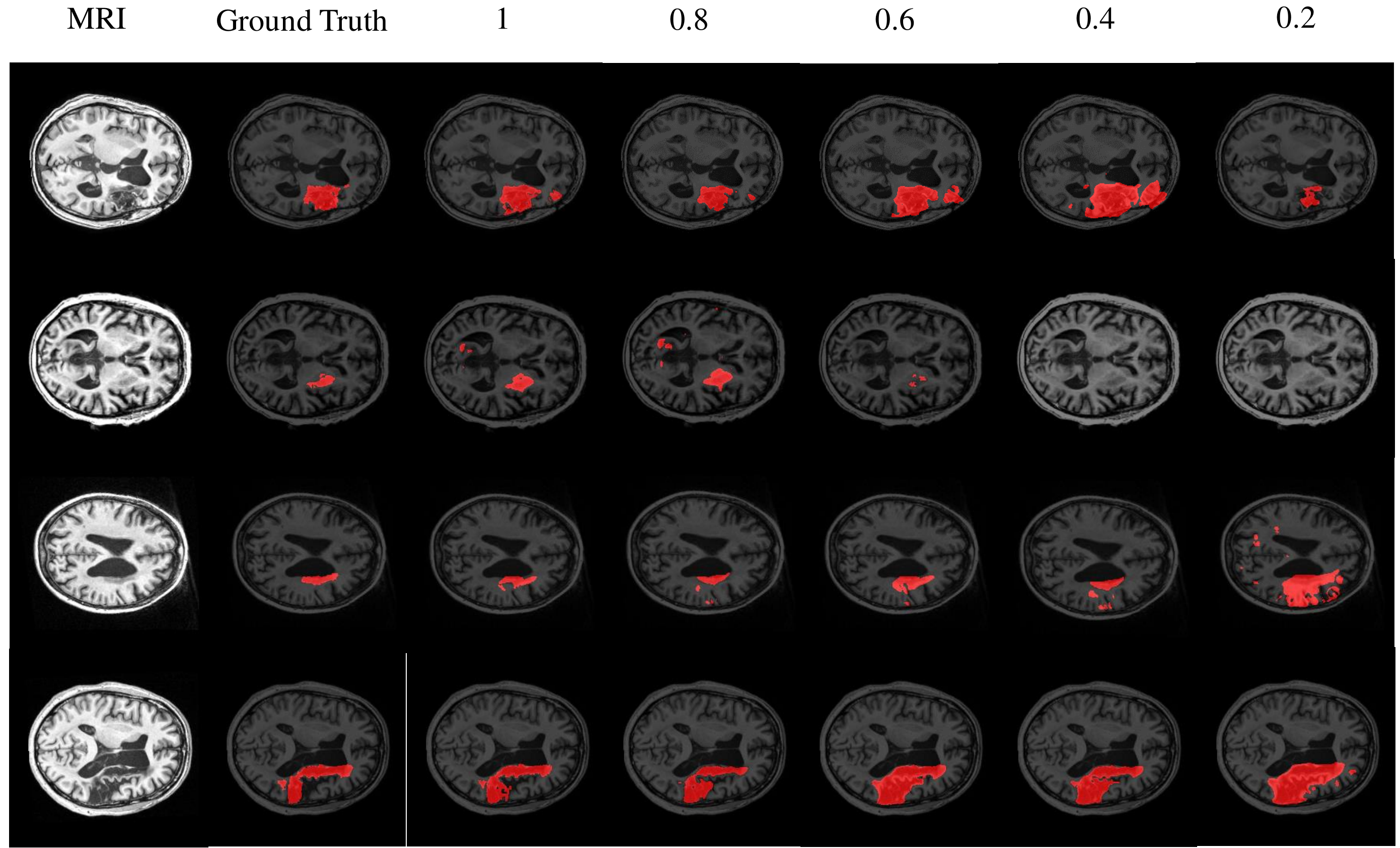}}
  \caption{{\color{blue}Semi-supervised segmentation results on ATLAS dataset.}}
\label{7_vs}
\end{figure*}

To further evaluate our model, we divide the testing set into 5 parts to draw the
box-plots.
Fig.\ref{3} shows the box-plots of Dice, Jaccard index, accuracy, sensitivity,
and specificity of different models. These results demonstrate that the performance of
our model is superior to other methods. Some details of the segmentation results are
shown in Fig.\ref{6_vs}. It can be inferred that our proposed CPGAN can segment the
brain stroke lesions in T1-weighted MR images very well. Our model performs well on
some fuzzy lesion boundaries and the confidential partition between stroke and
non-stroke regions.
It is very important to help specialists measure the stroke in stroke segmentation tasks.

\begin{figure*}[htp]
  \centerline{\includegraphics[width=\textwidth]{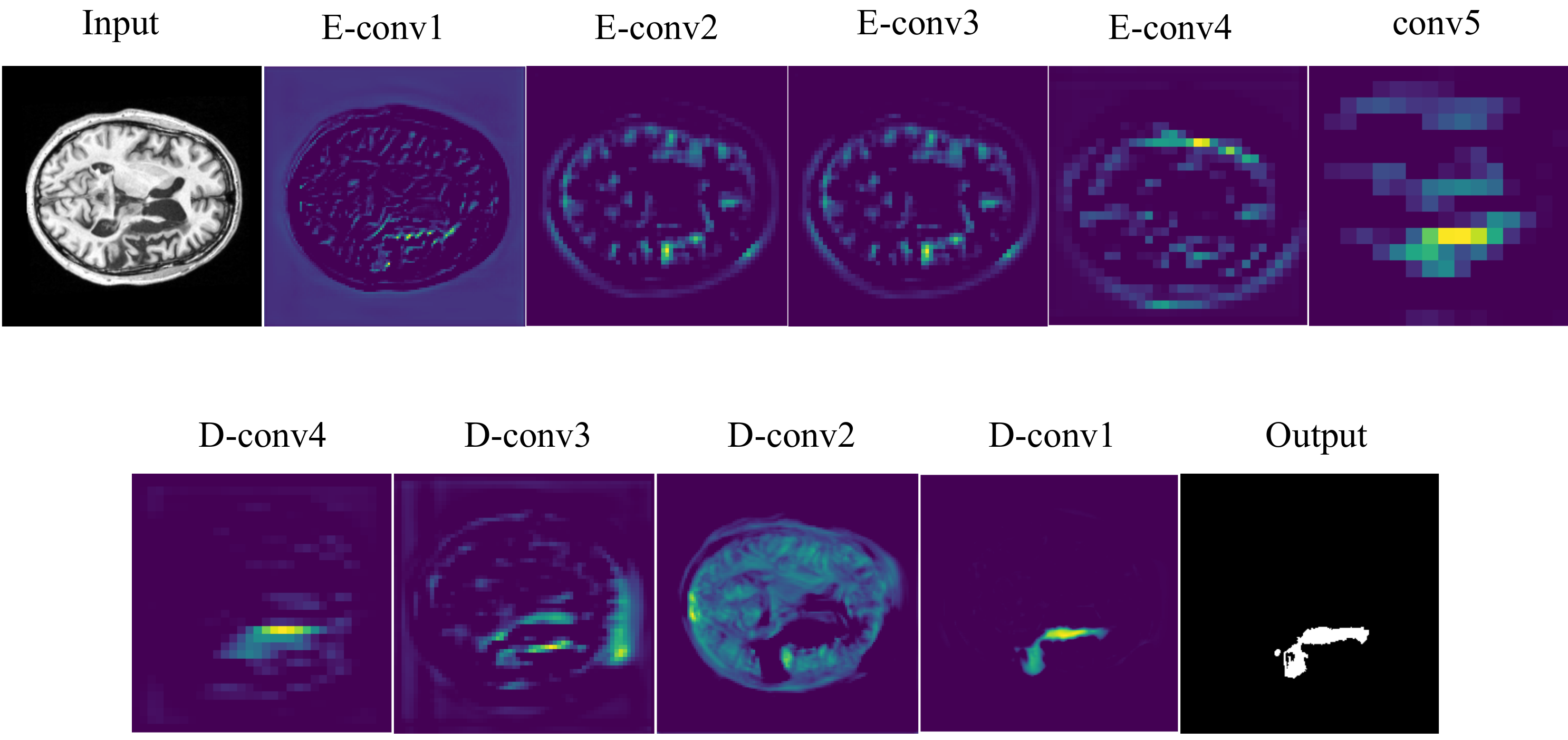}}
  \caption{Intermediate feature visualization of proposed CPGAN.
  }
\label{3feature}
\end{figure*}

To better understand what features our model has learned, each of layers in the CPGAN
are visualized in Fig.\ref{3feature}. From these feature visualizations, our proposed
CPGAN can capture the important pixel areas of stroke lesion in T1-weight MR images.
It demonstrates that our model has excellent ability in stroke lesion segmentation.

\subsubsection{Effectiveness of semi-supervised segmentation}

\begin{figure}[htp]
  \centerline{\includegraphics[width=0.5\textwidth]{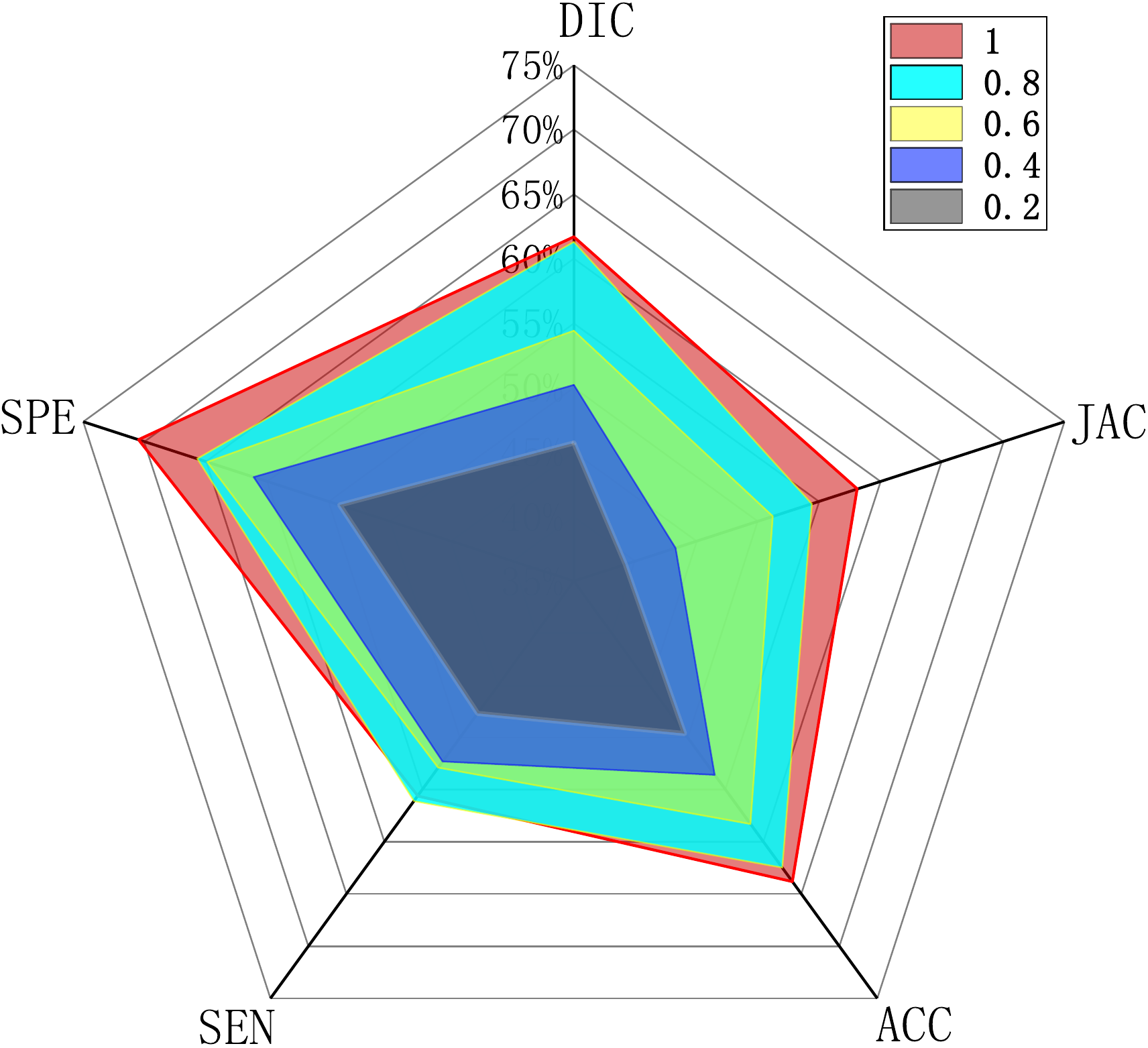}}
  \caption{Radar map of Dice, Jaccard index, Accuracy, Sensitivity and
  Specificity on the validation set.}
\label{k}
\end{figure}

To show the semi-supervised segmentation performance of CPGAN, 5 experiments are
designed and the number of labeled images are different for each set of experiments.
The rate of labeled images is set to 1, 0.8, 0.6, 0.4 and 0.2. As shown in the
Table\ref{tab3} and Fig.\ref{k}, our semi-supervised method performs better than
some full supervised method in different labeled/unlabeled data settings, which
demonstrates that our method effectively utilizes unlabeled data and proposed method is
beneficial to the performance gains. Only three-fifths of labeled images are used,
our proposed model performs better than X-net in Table with 0.055 improvement on
Jaccard index.
Furthermore, when the rate is set to two-fifths, the scores of Dice, Jaccard,
sensitivity and specificity are higher than U-net.
The comparison shows the effectiveness of our semi-supervised segmentation method
achieves the training effect of some full supervised methods.
Some semi-supervised segmentation results of different rates are shown in Fig.\ref{7_vs}.
It can be inferred that the proposed method presented stronger capability in semi-supervised
brain stroke lesion segmentation.

\begin{table}
  \renewcommand\arraystretch{1.5}
  \begin{center}
  \caption{Results of proposed method on the validation set under different rate of labeled images}
  \label{tab3}
  \normalsize
  \setlength{\tabcolsep}{2mm}{
  \begin{tabular}{ccccccc}
  \toprule
   Labeled/Full & Dic   & Jac   & Acc   & Sen   & Spe   \\\midrule
   1           & 0.617 & 0.581 & 0.638 & 0.556 & 0.705\\
  0.8          & 0.613 & 0.544 & 0.625 & 0.531 & 0.657 \\
  0.6          & 0.544 & 0.512 & 0.583 & 0.529 & 0.649 \\
  0.4          & 0.502 & 0.433 & 0.536 & 0.523 & 0.611 \\
  0.2          & 0.457 & 0.392 & 0.496 & 0.477 & 0.541 \\
  \bottomrule
  \end{tabular}}
  \end{center}
\end{table}

\section{Conclusion}
In this paper, we propose a novel semi-supervised segmentation method named CPGAN for
brain stroke lesion segmentation. The similarity connection module is adopted into the
segmentation network.
The effectiveness of the proposed similarity connection module is verified through
ablation study. This module can effectively improve the details of the lesion area of
segmentation by capturing long-range spatial information. The proposed assistant network
is pre-trained and shares the same architecture as discriminator.
The hyper-parameters of assistant network are fixed during the training.
The qualitative and quantitative experimental results demonstrate that representative
information of this network help discriminator mitigate the problem of discriminator
forgetting and improve performance of segmentation network.
The proposed consistent Perception strategy strategy is very useful to semi-supervised
segmentation.
Only two-fifths of labeled images are used, our proposed model performs better than
some other methods.
Results suggest that our method performs better in segmentation on the ATLAS dataset.
This method can also be extended to other medical image segmentation tasks.
In this work, we only experimented with one change of angle and found this transformation
strategy is very useful for semi-supervised segmentation.
In the future, we will explore the application of the consistent perception strategy
and add more transformations to improve the performance of semi-supervised segmentation.

\section*{Acknowledgments}
This work was supported by the National Natural Science Foundations of China under 
Grants 62172403 and 61872351, 
the International Science and Technology Cooperation Projects of Guangdong under 
Grant2019A050510030, the Distinguished Young Scholars Fund of Guangdong under 
Grant 2021B1515020019, 
the Excellent Young Scholars of Shenzhen under Grant RCYX20200714114641211 
and Shenzhen KeyBasic Research Project under Grants JCYJ20180507182506416 
and JCYJ20200109115641762.

 
 \bibliographystyle{IEEEtran}
 \bibliography{ref}

\end{document}